\documentclass[aps,groupedaddress]{revtex4}

\usepackage{amsmath,graphicx}

\begin{document}

%\preprint{}

\title{Rayleigh Scattering of Whispering Gallery Modes of Microspheres due to a Single Scatterer: Myths and Reality}

\author{L. Deych}
\author{J. Rubin}
\affiliation{Department of Physics, Queens College of the City University of New York (CUNY) Flushing, NY 11367}

\date{\today}

\begin{abstract}
The interaction of whispering gallery modes (WGM) of optical microresonators with subwavelength imperfections has been studied both experimentally and theoretically. This interaction is responsible for the formation of spectral doublets in place of single resonance peaks, and for degrading of Q-factors of the resonances. Within the currently accepted framework the spectral doublets are explained as a result of degeneracy removal of clockwise and counterclockwise WGMs due to their coupling caused by defect-induced backscattering, while the degrading of the Q-factor is described phenomenologically as an additional contribution to the overall decay rate of WGM due to coupling between WGM and radiative modes. Here we show that the existing understanding of this phenomenon is conceptually wrong and develop an exact theory of WGM interaction with a single defect, which provides a unified treatment for both aspects of this interaction explaining existing experiments and predicting new phenomena.
\end{abstract}

\pacs{}
%\keywords{photonic crystal, exciton, lattice, strong coupling, Bragg, quantum-well}

\maketitle

Elastic (with no change in frequency) scattering of light due to small (compared to wavelength) particles is one of the most fundamental and intensively studied optical phenomena. Its modern history began almost one hundred fifty years ago with the explanation of the blue color of sky in a series of papers by Lord Rayleigh\cite{Rayleigh}, where the now famous $1/\lambda^4$ cross section law, where $\lambda$ is the wavelength of light in vacuum, was derived.  Since then it has been customary to refer to processes of elastic interaction of light with subwavelength particles as Rayleigh scattering.
Besides providing us with beautiful blue skies and red sunsets, Rayleigh scattering is important for a large number of fundamental optical phenomena as well as for numerous applications.  Recent developments in optics and photonics have created new situations in which the manifestations of Rayleigh scattering are significantly modified. Particularly drastic modification of this process is expected when light is confined in all three dimensions inside optical microresonators in the form of whispering gallery modes (WGM)~\cite{VahalaNature2003}. Given the fundamental nature of this process
%The interaction between WGMs and various kinds of subwavlength imperfections of the confining structures is as fundamental for the optics of microcavities as standard Rayleigh scattering is for free propagating waves.
it is not surprising that it has attracted a significant amount of attention in recent years ~\cite{WeissOL95,LittleOL1997,Gorodetsky2000,KippenbergOL2002,BorselliOE2005}.

While whispering gallery modes can occur in various types of geometries~\cite{BoriskinaReview2006} we will focus on spherical microresonators. WGMs in this case correspond to Mie resonances~\cite{Mie1908} with ultra narrow widths, $\gamma_{ls}\ll \omega_{ls}$, where $\omega_{ls}$ is the frequency of the mode, and respectively high (up to $10^9$ for silica microspheres~\cite{BoriskinaReview2006}) Q-factors defined as $Q_{ls}=\omega_{ls}/\gamma_{ls}$. WGMs are characterized by polar and azimuthal indexes, $l$ and $m$, and a radial number $s$ determining, respectively, the angular and radial dependence of the fields in a spherical coordinate system centered at the sphere.
%Depending on polarization, transverse electric (TE) or transverse magnetic (TM), electromagnetic field of these modes is described by one of two types of vector spherical harmonics (VSH),
%\begin{equation}\label{eq:VSH_def}
%    \displaystyle{\mathbf {E}^{TE}_{lm}\propto \mathbf {M}_{lm}(k_{ls}r,\theta,\phi)}; \hskip 10pt \displaystyle{{\mathbf E}^{TM}_{lm}\propto \mathbf {N}_{lm}(k_{ls}r,\theta,\phi)},
%\end{equation}
%where variables $r$, $\theta$,and $\phi$ correspond to radial, polar and azimuthal angles of a particular spherical coordinate system centered at the particle, $k_{ls}=n\omega_{ls}/c$, where $c$ is the vacuum speed of light, and $n$ is either  refractive index of the sphere or $1$ depending on $r$ being inside or outside of the spherical particle respectively. Definitions of functions $\mathbf{M}_{lm}$ and $\mathbf{N}_{lm}$ are the same as in Ref.\onlinecite{stratton_book1941}, where polar $l$ and azimuthal $m$ ($|m|\le l$) numbers specify angular dependence of these functions. %Angular number $l$ determines eigenvalue,  $l(l+1)$, of the angular part of the %Laplace operator, while azimuthal number $m$ is the eigenvalue of the operator of rotation around polar axis $Z$ of the chosen coordinate system.
The resonance frequency $\omega_{ls}$ does not depend on the azimuthal number, which reflects the degeneracy of the resonances due to full spherical symmetry of the problem. %The latter is given by a spherical Bessel function inside %of the sphere, and by spherical Hankel function of the first kind outside of the sphere.
WGMs are also characterized by their the mode volume, which can be very different for modes with the same $l$ but different $m$. Modes with the smallest volume correspond to $|m|=l$, and $s=1$ in which case the field is concentrated mostly in the equatorial plane and at the surface of the sphere. Such modes are called fundamental (FM) and their interaction with defects is of the primary interest.

This interaction causes two observable effects: (i) formation of spectral doublets in place of a single peak, and (ii) reduction of  Q-factors of WGMs below theoretically predicted limits~\cite{BorselliOE2005,GrudininOC2006}. In the existing approaches these two effects are considered separately, even though they are two manifestations of the same phenomenon. The accepted explanation of the spectral doublets is based on the hypothesis likely first suggested by D.S. Weiss et al. in the following form: "We have observed that very high-Q Mie resonances in silica microspheres are split into doublets. This splitting is attributed to internal backscattering that couples the two degenerate whispering-gallery modes propagating in opposite directions along the sphere equator"~\cite{WeissOL95}. This idea of the defect-induced backscattering was further developed in subsequent publications~\cite{LittleOL1997,Gorodetsky2000,BorselliOE2005} and was claimed to be experimentally confirmed in Ref.~\cite{KippenbergOL2002}. In a recent paper by A. Mazzei, et al.~\cite{MazzeiPRL2007} double peak features in the spectra of microspheres were studied under conditions of controlled scattering, where the role of the defect was played by a tip of a near-field optical microscope. This work directly confirmed a connection between the interaction of WGMs with a \emph{single} defect and the formation of the spectral doublets.

The influence of defects on Q-factors of microresonators is usually prescribed to defect-induced coupling between WGMs and radiative modes and is taken into consideration phenomenologically by adding a "scattering" loss rate to the total losses of the resonator. In the case of disk resonators this rate was calculated in Ref.~\cite{BorselliOE2005} under the assumption that the surface roughness couples  WGMs with free space electromagnetic radiation.
\section{Results}\label{Results}
In this work we develop an \emph{ab initio} theory of interaction between WGMs of microspheres and a \emph{single} subwavelength defect, which treats both aspects of this interaction in a unified way. We show that the generally accepted picture of interaction between FM and defects is conceptually wrong: the interaction cannot be described in terms of "backscattering", and it does not result in coupling between counterpropagating "clockwise" (cw) and "counterclockwise" (ccw) FMs. The theory predicts that the observed spectral doublets are actually a part of a triplet of peaks, the third component of which has not yet been found. It also provides an accurate "from first principles" description of the broadening of the resonances.  Similar to the  Rayleigh scattering of propagating waves, the solution of the single-defect problem presented here constitutes the first fundamental step toward a theory of interaction between WGMs and multiple defects. When this interaction is small, which is usually the case, the single-defect solution can be directly used to make conclusions about the role of multiple scatterers. Since the results of Ref.~\cite{MazzeiPRL2007} directly confirm that the spectral features formed due to a single discrete scatterer are similar to those caused by distributed surface or volume disorder, the results presented here are relevant not only for interaction of WGMs with discrete non-interacting defects, but can also be used to understand effects due to continuously distributed nonuniformities.
\begin{figure}
   \includegraphics[width= .4\linewidth, angle=0]{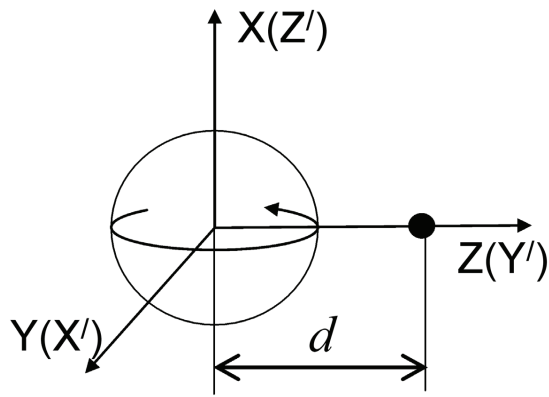}
  \caption{The coordinate systems used in calculations. The curved arrow schematically represents  the counterclockwise fundamental mode whose interaction with the defect (blackened small sphere) is under consideration.\label{fig:coordinates}}
\end{figure}

The main assumption of the theory is that the defect is small enough to be treated as a dipole. In this case the shape of the defect is irrelevant, and can be taken to be spherical.  In this way the problem is reduced to describing two electromagnetically coupled spheres of radii $R_0$ and $R_d\ll R_0$ with refractive indexes $n$ and $n_d$ respectively, whose centers are positioned at a distance $d$ from each other   (see Fig.~\ref{fig:coordinates}). We will also assume that the defect lies in the plane of the FM in order to maximize the strength of the interaction, although the theory developed here can readily be generalized for arbitrarily positioned defects. The goal is to find electromagnetic field induced by this system in the presence of an incident wave  (imitating a mode of a tapered fiber)  which, in the absence of the defect, would have excited a FM with given polar number $L$. The incident, $\mathbf{E}_{inc}$, and induced, $\mathbf{E}_s$, fields can be presented as linear combinations  of vector spherical harmonics (VSH) of the form:
\begin{eqnarray}
\mathbf{E_{inc}}&=&\sum_{m=-L}^{L}\eta_{L,m}\mathbf{N}_{L,m}(\mathbf{r}-\mathbf{r}_1)\label{eq:inc_ext}\\
\mathbf{E_{s}}&=&\sum_{i=1}^2\sum_{l=1}^\infty\sum_{m=-l}^l\left[a_{l,m}^{(i)}\mathbf{N}_{l,m}(\mathbf{r}-\mathbf{r}_i)+b_{l,m}^{(i)}\mathbf{M}_{l,m}(\mathbf{r}-\mathbf{r}_i)\right]\label{eq:scat_ext}
%\mathbf{E_{in}}=\sum_{i=1}^N\sum_{l,m}\left[c_{l,m}^{(i)}\mathbf{N}_{m,l}(\mathbf{r}-\mathbf{r}_i)+d_{l,m}^{(i)}\mathbf{M}_{m,l}(\mathbf{r}-\mathbf{r}_i)\right]\label{eq:intern_ext}.
\end{eqnarray}
where index $i$ enumerates the spheres ($i=2$ refers to the defect), $\mathbf{r_i}$ is a position vector of the center of $i-th$ sphere,  $\mathbf{M}_{l,m}$ and $\mathbf{N}_{l,m}$ are the VSH of TE and TM polarization respectively as defined in Ref.~\cite{stratton_book1941}, and $\eta_{L,m}$ are coefficients describing the TM incident wave of frequency $\omega$, which in the coordinate system $XYZ$ defined in Fig.~\ref{fig:coordinates} and used in all subsequent calculations have the following form
\begin{equation}\label{eq:FM}
\eta_{L,m}=(-1)^{\epsilon(L+m)}\frac{(-i)^L}{2^L}\displaystyle{\sqrt{\frac{(2L)!}{(L+m)!(L-m)!}}};
\hskip 4pt \epsilon=\left\{
\begin{array}{cc}
1 & \textit{cw FM} \\
0 & \textit{ccw FM}
\end{array}\right.
\end{equation}
(See details in Section~\ref{Methods}.) In order to find the induced field one needs to determine expansion coefficients $a_{l,m}^{(i)}$ and $b_{l,m}^{(i)}$. For simplicity we disregard the defect-induced coupling between TE and TM modes, and since the incident field is of TM polarization we set the TE coefficients $b_{l,m}^{(i)}=0$ and solve for coefficients $a_{l,m}^{(1,2)}$. The dipole approximation for the defect is introduced by setting $a_{l,m}^{(2)}=0$ for all $l>1$. The resulting system of equations for the scattering coefficients is solved exactly to yield:
\begin{eqnarray}
 a_{l,m}^{(1)}&=&\left\{ \begin{array}{lcc}
 \displaystyle{\frac{\eta_{L,m}}{[\alpha_L^{(1)}]^{-1}+(-1)^L\alpha_1^{(2)}A_{1,m}^{L,m}A_{L,m}^{1,m}+\alpha_1^{(2)}[\alpha_L^{(1)}]^{-1}\displaystyle{\sum_{\nu\ne L}(-1)^\nu\alpha_\nu^{(1)}A_{1,m}^{\nu ,m}A_{\nu ,m}^{1,m}}}} & l=L;|m|\le 1& (a) \\
 -\alpha_l^{(1)}\eta_{L,m}\displaystyle{\frac{(-1)^{L}\alpha_1^{(2)}\alpha_L^{(1)}A_{1,m}^{L,m}A_{l,m}^{1,m}}{1+\alpha_1^{(2)}\displaystyle{\sum_{\nu} (-1)^\nu\alpha_\nu^{(1)}A_{1,m}^{\nu ,m}A_{\nu,m}^{1,m}}}} & l\ne L;|m|\le 1& (b) \\
 \alpha_l^{(1)}\eta_{L,m}\delta_{l,L} & |m|>1 & (c)
 \end{array}\right.\label{eq:a1_coeff}\\
a_{l,m}^{(2)}&=&-\alpha_1^{(2)}\sum_\nu a_{\nu ,m}^{(1)}(-1)^\nu A_{1,m}^{\nu ,m}\label{eq:a2_coeff}
\end{eqnarray}
where  $\alpha^{(1,2)}$ are the single sphere scattering parameters for the main sphere and the defect respectively, defined in Eq.~\ref{eq:alpha_exact} and \ref{eq:alpha2_approx} of section~\ref{Methods}. The scattering parameter of the main sphere has poles at the complex-valued frequency of WGMs; we  assume that the frequency $\omega$ of the incident field is in the vicinity of the pole $\omega_L^{(0)}-i\Gamma_L^{(0)}$ corresponding to the frequency of our FM. Since the defect is assumed to be small so that $n_d\omega R_d/c \ll 1$, its scattering parameter does not have any poles of its own.

Parameters $A_{l,m}^{\nu ,m}$ in Eq.~\ref{eq:a1_coeff} and \ref{eq:a2_coeff} describe the electromagnetic interaction between the spheres and are called translation coefficients. They appear when a VSH defined in one coordinate system needs to be expressed in terms of VSH defined in a system with a shifted origin~\cite{SteinApplMath1961,CruzanApplMath1962,Mishchenko_book2002}. These coefficients depend on the translation vector $\mathbf{r}_1-\mathbf{r}_2$ and the coordinate system used to define VSHs. In the coordinate system $XYZ$  they are diagonal in terms of azimuthal number $m$, which reflects the fact that the polar axis $Z$ runs along the line connecting the centers of the spheres, thus preserving the axial symmetry of the two-sphere structure.
\begin{figure}
   \includegraphics[width= .4\linewidth, angle=0]{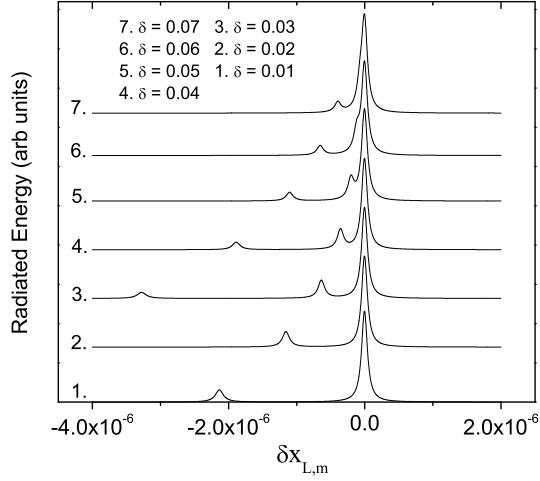}
  \caption{Radiated energy of the microsphere-defect system with varying distance parameter $\delta = (d-R_0-R_d)/R_0$.\label{fig:scatter_energy}}
\end{figure}
Eqs.~\ref{eq:a1_coeff} and \ref{eq:a2_coeff} contain all the information about the electromagnetic field of the sphere-defect system. First of all, Eq.~\ref{eq:a1_coeff}c shows that components of the initial FM with $|m|>1$ are not effected by the defect. Formally, this result is a consequence of the translation coefficients $A_{l,m}^{\nu ,\mu}$ being equal to zero when either $m$ or $\mu$ exceeds either $\nu$ or $l$~\cite{Mishchenko_book2002}. Physically, this result reflects the simple fact that a dipole can only produce a field with $l=1$ and $m=0,\pm 1$. Since in the $XYZ$ coordinate system $m$ remains a conserving quantity even in the presence of a defect there will be no coupling between the field of defect and WGMs with $|m|>1$. Thus, the expansion coefficients $a_{l,m}^{(1)}$ with $l=L$ and $|m|>1$ will produce a resonance at the original single-sphere frequency, which we will characterize by  size parameter $x_L^{(0)}=\omega_L^{(0)}R_0/c=k_L^{(0)}R_0$. The width of this resonance, which can be described by a dimensionless parameter $\gamma_L^{(0)}=\Gamma_L^{(0)}R_0/c$ is also not affected by the defect. Eq.\ref{eq:a1_coeff}a, on the other hand,  shows that $|m|\le 1$ components of the FM do interact with the defect, and that this interaction results in appearance of new resonance frequencies determined by poles of this expression. From the sum over $\nu$ we singled out a term with $\nu=L$, which in the frequency range around $\omega_L^{(0)}$ gives the biggest contribution to the shift of the new poles from their single-sphere value. Discarding the rest of the sum (the resonance approximation) we can obtain an analytical expression for the positions of the new  poles
\begin{equation}\label{eq:res_pos}
\begin{split}
&x_{L,m}=x_L^{(0)}(1+\delta x_{L,m}) -i\gamma_L^{(0)}(1+\delta\gamma_{L,m})\\
\delta x_{L,m} = &-\gamma_L^{(0)}\frac{f_{L,m}\left(k_L^{(0)}d\right)}{(2L+1)R_0^2d}p[x_L^{(0)}]; \hskip 4pt
\delta\gamma_{L,m} = \frac{2}{3}\frac{f_{L,m}\left(k_L^{(0)}d\right)}{(2L+1)R_0^5d}p^2[x_L^{(0)}]^5
\end{split}
\end{equation}
where
\begin{equation}\label{eq:polar_unnorm}
p=\frac{n_d^2-1}{n_d^2+2}R_d^3
\end{equation}
is the standard  polarizability of a small dielectric sphere, and function $f_{L,m}\left(k_L^{(0)}d\right)$ for $m=0,\pm 1$ is defined as
\begin{equation}\label{eq:f_funct}
f_{L,m}(kd)=\left[{(-1)^m}\sqrt{\frac{(L+1)(L+m^2)}{1+m^2}}g_{L-1}(kd)+\sqrt{L(L+1)(1-m^2)+L^2\frac{m^2}{2}}g_{L+1}(kd)\right]^2
\end{equation}
with
\begin{equation}
g_L(kd) = {\frac{1}{\sqrt{\rho\xi}}}{e^{\xi(atanh{\rho}-\rho)}} \  \  ;  \  \
\rho = \sqrt{1-\left(\frac{kd}{\xi}\right)^2} \  \ ; \  \
\xi = L + \frac{1}{2} \nonumber
\end{equation}
where we used an asymptotic form of the Hankel function valid for $l\gg kd$. Eq.~\ref{eq:res_pos} predicts two new resonances, in addition to the original single sphere resonance, one for $m=0$ and another for $m=\pm 1$, both red shifted with respect to the initial frequency. Function $f_{Lm}(kd)$ specifies the dependence of the resonance frequencies on $m$ and distance $d$. The later is determined by the exponential decay of the spherical Hankel functions outside of the main sphere, which reflects the evanescent nature of the interaction with the defect. Thus, our theory predicts the existence of a triplet of peaks rather than the doublet expected in the current cw-ccw coupling picture. To verify this result we carried out numerical calculations of the frequency dependence of the energy emitted by the main sphere given by the standard expression $\sum_l(2l+1)|a_{l,m}|^2$~\cite{stratton_book1941}, using complete Eq.~\ref{eq:a1_coeff}. For these calculations we choose $L=39$ and take into account enough coefficients $a_{l,m}$ and terms in the sum over $\nu$ to ensure convergence of the procedure, which was achieved with $1\le l\le 50$ and $\nu\le 50$. The results of these calculations are shown in Fig.~\ref{fig:scatter_energy} for different distances between the defect and the sphere so that one can see how the peaks shift toward the single-sphere resonance with increasing $d$ and eventually merge with it.   There are indeed three peaks, which are not seen at curves 1 and 2 because the third peak on these curves is out of the range of the figure. The $m=0$ resonance is shifted further from $x_L^{(0)}$ and is weaker than the $m=\pm 1$ resonance, making it more difficult for experimental identification. We suggest, therefore, that the experimentally observed spectral doublets correspond to the original single-sphere resonance and the $m=\pm 1$ resonance introduced by the defect.
\begin{figure}
   \includegraphics[width=.4\linewidth,angle=0]{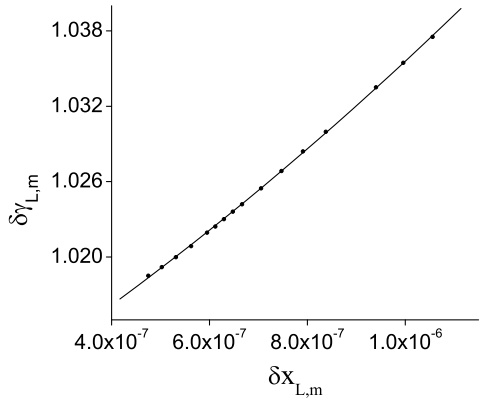}
  \caption{Relative broadening versus frequency shift of the $|m| = 1$ resonance revealed through varying distance $d$. The points are obtained from the spectra computed at different $d$, while the line represents a fit with a quadratic polynomial\label{fig:shift_broad}}
\end{figure}

The validity of the resonance approximation depends on the convergence of the sum $\sum_{\nu\ne L}(-1)^\nu\alpha_\nu^{(1)}A_{1,m}^{\nu,m}A_{\nu,m}^{1,m}$ appearing in the denominator of Eq.~\ref{eq:a1_coeff}a. In the limit $\nu\rightarrow\infty$ we find
\begin{equation}\label{eq:asympt}
    (-1)^{\nu+1}\alpha_\nu^{(1)}A_{1,m}^{\nu,m}A_{\nu,m}^{1,m}\asymp \frac{p^2}{(kd)^3}\left(\frac{R_0}{d}\right)^{2\nu+1}
\end{equation}
which, given that $R_0/d<1$, proves the convergency of the sum. However, in the case of small defects positioned close to the surface of the sphere, $R_0/d$ differs from unity by a small amount and the convergence of the sum is  slow. In this case the terms with $p\gg L$ become important and should be taken into account. It can be shown that incorporating these terms does not change the form of Eq.~\ref{eq:res_pos}  but renormalizes the polarizability, which becomes:
\begin{equation}\label{eq:renorm_p}
\tilde{p}=p\left[1+\left(1-\frac{m^2}{2}\right)\frac{dR_d^3(R_0^2+d^2)}{(kd)^3(d^2-R_0^2)^3}\frac{n^2-1}{n^2+2}\right]^{-1}
\end{equation}
The renormalized polarizability acquires dependence on the distance $d$ between the defect and the sphere, thereby affecting the relation between the frequency shift and the broadening of the defect-induced resonances which is revealed through variation of $d$. Indeed, in the absence of the renormalization both these quantities decrease with $d$ by the same factor determined by the function $f_{L,m}$, resulting in $\delta x_{L,m}\propto \delta\gamma_{L,m}$. The renormalized polarizability, however, also changes with distance.  Since the frequency shift of the resonance is linear in $p$ while the broadening is $quadratic$, this effect must result in deviations from this linear dependence. In order to confirm this conclusion we used numerical spectra obtained for different distances in order to plot $\delta x_{L,m}$ versus $\delta\gamma_{L,m}$ for the $|m|=1$ resonance. The obtained data shown in Fig.~\ref{fig:shift_broad} are found to be better fit by a quadratic rather than a linear function confirming this conclusion.

\begin{figure}
   \includegraphics[width=.8\linewidth,angle=0]{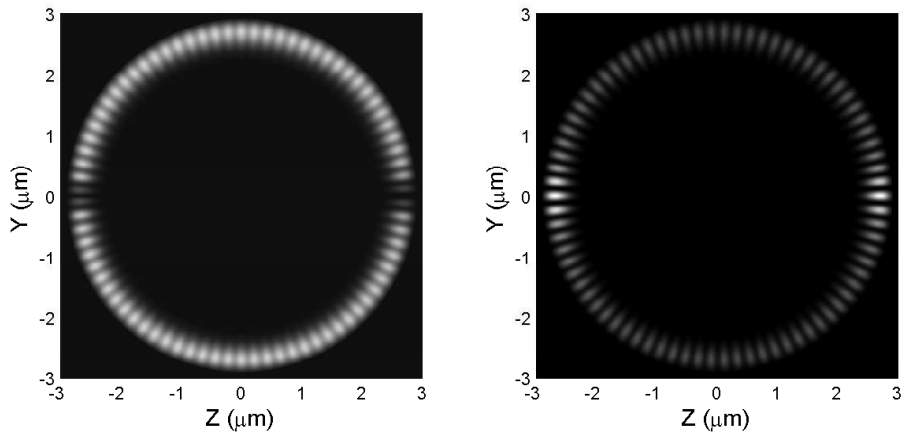}
  \caption{Internal field intensity of the microsphere in the YZ plane at the frequency of the standard Mie resonance (left) and the defect induced resonance (right).\label{fig:field}}
\end{figure}

With scattering coefficients known we can also compute the internal field inside the main sphere. Fig.~\ref{fig:field} shows the variation of the field in the $YZ$ plane of the $XYZ$ system (which corresponds to the plane of the FM) obtained by varying radial and polar coordinates at the azimuthal angles $\phi=\pi/2$ and $\phi=3\pi/2$. The computed field profile for the defect-induced peak demonstrates $2L$ oscillations and a drastic increase in intensity in the vicinity of the defect. At the frequency of a single-sphere resonance the situation is reversed: $2L$ oscillations, which are phase shifted compared to the defect-induced resonance are accompanied by a significant decrease in the field's intensity in the defect's proximity.

Finally Eq.~\ref{eq:a1_coeff}b describes coupling of the FM to other WGMs,  most important of which are terms with $l<L$. There are two reasons for this. First, modes with lower $l$ and higher radial numbers can spectrally overlap with the $l=L$, $s=1$ mode~\cite{DeychRoslyakPRE2006} and thus have a large effect on the field distribution. Second, these modes usually have lower $Q$-factors and contribute more significantly to the radiation losses of the system. Note that $m$-components with $m>1$, which are responsible for the resonance at the single-sphere frequency do not couple to any other modes, so that this resonance is not affected by the coupling to the low-Q WGMs. The effects of coupling to these modes  at the frequency of the defect-induced peak is shown in Fig.~\ref{fig:radiative_coupling}, where we plot the spectrum of the radiated energy in its vicinity with and without contributions of terms with $l\ne L$.
\section{Discussion}\label{Discussion}
The theory presented in the paper gives a complete picture of the interaction between WGMs and a single defect based on fundamental principles with no \emph{ad hoc} assumptions, and replaces the currently accepted paradigm, which is proved to be inadequate. The  results obtained on the basis of this theory show that a number of experiments previously "explained" within the cw-ccw coupling picture such as the backscattering experiment of Ref.~\cite{KippenbergOL2002} must be reinterpreted. They also provide a natural explanation to other experimental results that the current paradigm was not able to explain in addition to predicting new effects that await experimental confirmation.

To begin with, the developed theory gives a natural explanation of an asymmetry between the two peaks of a doublet, which was seen in all experimental observations of this effect, but most clearly in Ref.~\cite{MazzeiPRL2007}. Since, according to our calculations, the higher frequency component of the doublet corresponds to a single sphere resonance unaffected by the defect, this peak is supposed to be narrower than its counterpart and not to shift with the change in the position of the defect.  This behavior is in complete agreement with observations of Ref.~\cite{MazzeiPRL2007}. In addition, our theory predicts the existence of the third peak, which is, however, weaker than the other two making its experimental observation more difficult. It should be noted, however, that since within the prevailing paradigm the presence of the third peak was not expected, it is possible that more careful experimental observations will reveal its presence. Eq.~\ref{eq:res_pos} also predicts quite specific dependence of the peak's position and its broadening on the position of the defect and the polar number $L$ and frequency $\omega_L^{(0)}$, which also can be verified experimentally.

\begin{figure}
   \includegraphics[width=.5\linewidth,angle=0]{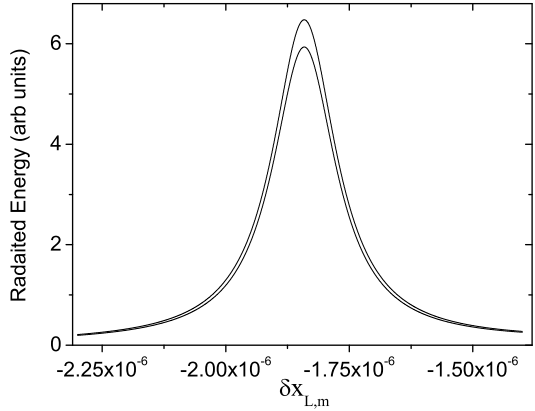}
  \caption{Radiated energy with (the curve with the higher peak) and without contributions from $l\neq L$.\label{fig:radiative_coupling}}
\end{figure}

We also found that while the main contribution to the width of the resonance comes from coupling to the dipole field of the defect, there is also an additional contribution from coupling to lower Q WGMs of the main sphere. This contribution is emphasized in Fig.~\ref{fig:radiative_coupling}, where we compare defect-induced resonance with and without terms with $l\ne L$. One can see that these terms make the resonance wider while increasing its height. This effect, which cannot be described by simply adding an additional loss term to the Q-factor, is relatively small for a single defect case, but can be expected to become more significant with multiple defects.
%It is interesting to compare our Eq.~\ref{eq:res_pos} with similar results obtained on the base of heuristic arguments in Ref.\cite{MazzeiPRL2007}. Besides the obvious facts that heuristic arguments cannot predict the dependence of the resonance frequency and its width on the position of defect  and the polar number of the FM, the dependence on the frequency of the FM predicted in Ref.\cite{MazzeiPRL2007} is striking contrast with our results. Our calculations give quadratic dependence on $x_L^{(0)}$ of the frequency shift of the defect induced resonance and fifth power dependence for the broadening instead of respective linear dependence and fourth power dependence predicted by Mazzei et al. The origin of this difference can be traced to difference between the far field $1/x$ asymptotic  of $h_l(x)$ and its $1/\sqrt{x}$ behavior in the near field region, which was used in deriving Eq.~\ref{eq:res_pos}.  Calculations of Mazzei et al. were based on incorporating scattering properties of propagating waves into the WGM problem, which is equivalent to considering WGM in the far field region.

Another important effect predicted in our theory is the position dependent renormalization of the polarizability of the defect. This effect results in a deviation from the linear dependence  between the frequency shift of the defect-induced resonance and its broadening, which was recently observed in Ref.~\cite{MazzeiPRL2007}. The authors of that paper suggested that the position dependence of the polarizability could account for this finding, but were unable to explain its origin. In our theory this dependence appears naturally as a result of coupling between the defect and WGMs with high polar numbers $l$.

An important characteristic of the interaction between the FM and the defect is the resulting distribution of the electromagnetic field along the surface of the sphere. The backscattering paradigm predicts the formation of standing waves with $2L$ oscillations of the field's intensity along the circumference of the FM. These waves are assumed to be due to interference of cw and ccw modes and are described by  either $\sin$- or $\cos$-like behavior, depending on which component of the doublet is considered. Using Eq.~\ref{eq:FM} we can see that in $XYZ$ coordinate system used in our calculations these standing waves should be described as
\begin{equation}
\mathbf{E_{sw}}=\sum_{m=-L}^{L}\frac{(-i)^L}{2^L}\displaystyle{\sqrt{\frac{(2L)!}{(L+m)!(L-m)!}}}\mathbf{N}_{L,m}(\mathbf{r}-\mathbf{r}_1)\label{eq:stand_wave}
\end{equation}
where for the symmetric combination of the cw and ccw modes $m$ takes on only even values for even $L$, and odd values for odd $L$; for antisymmetric combination the situation is reversed. Our results show that for both the defect-induced and single sphere resonance there are indeed $2L$ oscillations, which, however, do not have the form prescribed by Eq.~\ref{eq:stand_wave}. The field distribution at the frequency of the defect-induced peak is explained by the fact that the field at this frequency is mainly comprised of the $m$-components with $|m|=1$. The field of these WGMs is characterized by $L-|m|+1=L$ oscillations for $\theta$ changing between $0,\pi$ giving their total number equal to $2L$. These modes are also characterized by the enhancement of the field in the vicinity of $\theta=0$, which explains a drastic rise in the intensity around the location of the defect. The field distribution at the single sphere resonance can be understood by noting that this field is comprised of modes with $|m|>1$, which when added to the remaining $|m|\le 1$ components, would have produced a flat distribution of the intensity. Therefore, removal of these components obviously results in the decrease of the field around the defect and phase shifted oscillations elsewhere. The presence of these oscillations of the field's intensity demonstrate that one can explain experimental results of Ref.~\cite{KippenbergOL2002} without reliance on the "backscattering" paradigm.

Finally, we should comment on the relation between our results and the multi-defect problem. In the approximation of non-interacting defects, the field in the presence of multiple defects can be found as simple sum of fields due to each defect separately. Therefore, the generalization of these results would include finding the field distribution for a generic position of the defect relative to the plane of the FM. Since interaction with each defect can be considered independently we can always analyze it in a coordinate system with polar axis passing through the centers of the defect and the main sphere. Then, we can repeat all our calculations with only one adjustment: Eq.~\ref{eq:FM} needs to be generalized to incorporate an arbitrary inclination of the plane of the FM with respect to the coordinate axes. It is clear, therefore, that while the resulting field distribution and the heights of the resonance peaks will be different from the ones obtained here, the real and imaginary parts of the resonance frequencies will remain the same as in the single-defect case as long as all defects are identical~\footnote{ A size dispersion of the defects will result in additional inhomogeneous broadening of the resonances, but provided its statistical distribution is uniform would not change the position of the peak}. It is also important  to note that these quantities do not depend on the type of the initial FM, cw or ccw, which means that even in the presence of a defect or multiple defects there are still two  modes originating from cw or ccw initial FMs, which remain degenerate. This conclusion has a number of far reaching implications and dispels another important myth of the cw-ccw coupling paradigm, which suggests that in the presence of the defects each resonant peak corresponds to a single non-degenerate mode.
\section{Methods}\label{Methods}
Results presented in this paper were obtained with the help of a number of qualitative and quantitative methods. We will start with the approach based on symmetry arguments, which provides a qualitative explanation of many of the findings.
\subsection{Symmetry considerations and choice of the system of coordinates}
 The logic behind the ccw-cw coupling paradigm is based upon an implicit assumption that the degeneracy between cw and ccw modes, which persists even in the absence of the complete spherical symmetry, is due to the rotational symmetry with respect to the axis perpendicular to the plane of the FM. Indeed, if this were the case, then any defect would have violated this symmetry, lifting the cw-ccw degeneracy and resulting in the spectral doublet. This assumption, however, is not correct, which can be immediately seen if one recalls that the group of rotations about a single axis is Abelian and, therefore, can only have one-dimensional representations. This means that axial symmetry alone cannot explain the cw-ccw degeneracy and one needs to invoke an additional symmetry such as inversion with respect to the azimuthal angle $\phi$. In the case of a single sphere the inversion symmetry is "hidden" behind the more powerful spherical symmetry, but in the system with a defect it starts playing a significant role. Indeed, two interacting spheres, described in $X^\prime Y^\prime Z^\prime$ coordinates (Fig.~\ref{fig:coordinates}), which is not consistent even with the remaining axial symmetry of the system, still exhibits a symmetry with respect to replacement $\phi \rightarrow -\phi$ if the $X$-axis of that coordinate system is chosen  along the line connecting the centers of the spheres. Since this is the symmetry ultimately responsible for the cw-ccw degeneracy and it is not destroyed even in the system with the defect, the alleged coupling between cw and ccw modes cannot take place.

The $X^\prime Y^\prime Z^\prime$ coordinate system is convenient to describe the FM excited in the main sphere shown  in Fig.~\ref{fig:coordinates}. In this case the respective field can be presented as a single vector spherical harmonic of $TE$ or $TM$ polarization~\cite{stratton_book1941}, meaning that the expansion coefficients of Eq.~\ref{eq:inc_ext} takes a simple form $\eta_{L,m}=\delta_{L,m}$ instead of those given in Eq.~\ref{eq:FM}.   However, in the two-sphere problem this coordinate system is not consistent with the symmetry of the configuration, therefore, it is more convenient, following Ref.~\cite{MiyazakiPRB2000,deychPRA2008}, to switch to a coordinate system with polar axis directed along the line connecting the centers of the spheres (designated as $XYZ$ in Fig.\ref{fig:coordinates}).  It is important to realize, however,~\cite{deychPRA2008} that the field of this FM cannot be presented as a single VSH in the spherical coordinates based on the coordinate system $XYZ$. To obtain such representation we notice that this system is obtained from $X^\prime Y^\prime Z^\prime$ system by means of a rotation characterized by Euler angles $\alpha=\pi/2$, $\beta=\pi/2$, $\gamma=0$, where we are following notations from Ref.~\cite{Mishchenko_book2002}. Now using the transformation properties of VSH~\cite{Mishchenko_book2002} we obtain the representation of the FM in the $XYZ$-based spherical coordinates given in Eqs.~\ref{eq:inc_ext} and ~\ref{eq:FM}.

The $XYZ$ coordinate system reflects the presence of the axial symmetry of our configuration with respect to rotation about the axis connecting the centers of the spheres. Because of this symmetry, even though the modes of the sphere with the defect can no longer be classified according to the polar number, $l$, they still can be characterized by azimuthal number, $m$. Respectively, each of the $m$-components comprising the FM interacts with the defect independently making the analysis of the interaction simpler.
\subsection{Multi-sphere Mie theory}
In order to find the expansion coefficients of the induced field introduced in Eq.~\ref{eq:scat_ext} we use the standard multi-sphere
Mie theory~\cite{MiyazakiPRB2000,Mishchenko_book2002,FullerApplOpt1991}.
In this approach, the field outside of the spheres is separated into incident field given by Eq.~\ref{eq:inc_ext} and the induced field given by Eq.\ref{eq:scat_ext}. In addition the field inside the spheres is also presented as a linear combination of VSHs centered at each sphere
   \begin{equation}
%\mathbf{E_{inc}}=\sum_{i=1}^N\sum_{l,m}\left[\zeta_{l,m}^{(i)}\mathbf{N}_{m,l}(\mathbf{r}-\mathbf{r}_i)+\eta_{l,m}^{(i)}\mathbf{M}_{m,l}(\mathbf{r}-\mathbf{r}_i)\right]\label{eq:inc_ext}\\
%\mathbf{E_{s}}=\sum_{i=1}^N\sum_{l,m}\left[a_{l,m}^{(i)}\mathbf{N}_{m,l}(\mathbf{r}-\mathbf{r}_i)+b_{l,m}^{(i)}\mathbf{M}_{m,l}(\mathbf{r}-\mathbf{r}_i)\right]\label{eq:scat_ext}\\
\mathbf{E_{in}^{(i)}}=\sum_{l,m}\left[c_{l,m}^{(i)}\mathbf{N}_{l,m}(\mathbf{r}-\mathbf{r}_i)+d_{l,m}^{(i)}\mathbf{M}_{l,m}(\mathbf{r}-\mathbf{r}_i)\right]\label{eq:intern_ext}.
\end{equation}
In order to apply Maxwell boundary conditions on the surface on a sphere $i$ the  VSHs centered at different spheres must be rewritten in the coordinate system translated to the center of the $i$-th sphere. This is accomplished with the help of the addition theorem for the vector
spherical harmonics~\cite{CruzanApplMath1962,SteinApplMath1961}, which introduces translation coefficients $A_{l,m}
^{l^\prime,m^\prime}\left(x,\mathbf{r}_j-\mathbf{r}_i\right)$ and $B_{l,m}
^{l^\prime,m^\prime}\left(x,\mathbf{r}_j-\mathbf{r}_i\right)$, and which allows one to
derive a system of equations relating expansion coefficients
$a_{l,m}^{(i)}$ and $b_{l,m}^{(i)}$ to the coefficients of the incident
field $\eta_{l,m}^{(i)}$:
\begin{eqnarray}
a_{l,m}^{(i)}&=&\alpha_{l}^{(i)}\left\{\eta_{l,m}^{(i)}+\sum\limits_{j\neq i}\sum\limits_{l^\prime,m^\prime}\left[a_{l^\prime,m^\prime}^{(j)}A_{l,m}
^{l^\prime,m^\prime}\left(x,\mathbf{r}_j-\mathbf{r}_i\right)  +b_{l^\prime,m^\prime}^{(j)}B_{l,m}^{l^\prime,m^\prime}\left(x,\mathbf{r}_j-\mathbf{r}_i\right)\right]\right\}\label{eq:a_coeff_expan} \\
b_{l,m}^{(i)}&=&\zeta_{l}^{(i)}\sum\limits_{j\neq
i}\sum\limits_{l^\prime,m^\prime}\left[b_{l^\prime,m^\prime}^{(j)}A_{l,m}
^{l^\prime,m^\prime}\left(x,\mathbf{r}_j-\mathbf{r}_i\right)
+a_{l^\prime,m^\prime}^{(j)}B_{l,m}^{l^\prime,m^\prime}\left(x,\mathbf{r}_j-\mathbf{r}_i\right)\right]\label{eq:b_coeff_expan}
\end{eqnarray}
where $\alpha_l^{(i)}$ and $\zeta_l^{(i)}$ are single sphere Mie
scattering parameters for $TM$ and $TE$ polarizations
respectively. For the TM polarization this parameter, defined in terms of
dimensionless frequency parameter $x=R_0\omega/c$, is given by the well-known expression~\cite{Mishchenko_book2002}
\begin{equation}\label{eq:alpha_exact}
\alpha^{(1)}_l= - \frac{{j_l(x){\frac{d}{dx}}[xj_l(nx)]-{{n}^2}j_l(nx){\frac{d}{dx}}[xj_l(x)]}}{h_l(x){\frac{d}{dx}}[xj_l(nx)]-{{n}^2}j_l(nx){\frac{d}{dx}}[xh_l(x)]}
\end{equation}
where $j_l(x)$ and $h_l(x)$ are Bessel and Hankel functions respectively.  For the defect we only need $l=1$ and if $n_dx_d\ll 1$, where $x_d=xR_d/R_0$ the scattering parameter $\alpha^{(2)}_1$ does not have any poles and can be approximated as
\begin{equation}\label{eq:alpha2_approx}
\alpha^{(2)}_1\approx -\left(1+i\frac{3}{2}\frac{1}{p(n_dx_d)^3}\right)^{-1}
\end{equation}
Explicit expressions for translational coefficients
$A_{l,m}^{l^\prime,m^\prime}\left(\mathbf{r}_j-\mathbf{r}_i\right)$
and
$B_{l,m}^{l^\prime,m^\prime}\left(\mathbf{r}_j-\mathbf{r}_i\right)$,
which describe optical coupling between the spheres via modes of the
same or different polarizations respectively, can be found, for
instance in
Ref.~\cite{Mishchenko_book2002,FullerApplOpt1991,MiyazakiPRB2000}.
Important property of the translation coefficients is that they take a diagonal form in $m$  if the translation vector is parallel to the polar axis of the coordinate system used to define spherical coordinates. This significantly simplifies the equations for expansion coefficients eliminating summation over the azimuthal number and decoupling equations for coefficients with different $m$. We take advantage of this property by working in the $XYZ$ coordinate system of Fig.~\ref{fig:coordinates}.

For WGM with $l\gg 1$ the cross-polarization translation coefficients are usually much smaller than their same-polarization counterparts. Since we assumed incident wave to be of TM polarization we can set $b_{l,m}=0$ in Eq.~\ref{eq:a_coeff_expan} obtaining as a result a closed system of equation for coefficients $a_{l,m}$.
\subsection{Dipole approximation}
The field of the dipole is described by VSHs with $l=1$. Therefore, we introduce the dipole approximation by assuming that $a_{l,m}^{(2)}=0$ for $l>1$. This reduces the system of Eq.~\ref{eq:a_coeff_expan} to a simpler form
\begin{eqnarray}
a_{l,m}^{(1)}&=&\alpha_{l}^{(1)}\left\{\eta_{l,m}^{(i)}+a_{1,m}^{(2)}A_{l,m}^{1,m}\left(\mathbf{r}_1-\mathbf{r}_2\right)\right\}\label{eq:a1_coeff_simpl} \\
a_{1,m}^{(2)}&=&\alpha_{l}^{(2)}\sum\limits_{\nu}(-1)^{1+\nu}a_{\nu,m}^{(1)}A_{1,m}^{\nu,m}\left(\mathbf{r}_1-\mathbf{r}_2\right)\label{eq:a2_coeff_simpl}
\end{eqnarray}
which can be solved exactly by multiplying Eq.~\ref{eq:a1_coeff_simpl} by $(-1)^{1+l}A_{1,m}^{l,m}$ and summing over $l$. Substituting Eq.~\ref{eq:a2_coeff_simpl} into the resulting expression, we obtain a closed equation for the quantity $\sum_{\nu}(-1)^{1+\nu}a_{\nu,m}^{(1)}A_{1,m}^{\nu,m}$, which can be easily solved. As a result we arrive at Eq.~\ref{eq:a1_coeff} and \ref{eq:a2_coeff} of Section~\ref{Results}.
\begin{acknowledgments}
Financial support by AFOSR via grant F49620-02-1-0305, as well as support by PSC-CUNY grants is acknowledged.
\end{acknowledgments}

\end{document}